\documentclass[useAMS,usenatbib]{mn2e}
\usepackage{times}
\usepackage{graphicx}

\newcommand{\xmm}{{\it XMM~\/}}
\newcommand{\xmmn}{{\it XMM-Newton~\/}}
\newcommand{\asca}{{\it ASCA~\/}}
\newcommand{\chandra}{{\it Chandra~\/}}

\newcommand{\rosat}{{\it ROSAT~\/}}

\def\Msun{\hbox{$\rm M_{\odot}~$}}
\def\ergcms{{\rm ~erg~cm^{-2}~s^{-1}}}

\def\ergsec{{\rm ~erg~s^{-1}}}
\def\chisq{{$\chi^{2}$}}

\def\atpcm{{\rm ~atoms~cm^{-2}}}

\def\dg{^{\circ}}
\def\H0{{\rm ~km~s^{-1}~Mpc^{-1}}}

\def\eg{{\it e.g.,~\/}}

\def\ie{{\it i.e.,~\/}}

\def\la{\mathrel{\hbox{\rlap{\hbox{\lower4pt\hbox{$\sim$}}}{\raise2pt\hbox{$<$}}}}}
\def\ga{\mathrel{\hbox{\rlap{\hbox{\lower4pt\hbox{$\sim$}}}{\raise2pt\hbox{$>$}}}}}
\def\d25{D$_{25}$}
\def\nh{{$N_{H}$}}

\def\deg{\hbox{$^\circ$~\/}}
\def\arcm{\hbox{$^\prime$~\/}}
\def\arcs{\hbox{$^{\prime\prime}$~\/}}

\begin{document}
\title[A dipping black-hole X-ray binary 
candidate] {A dipping black-hole X-ray binary 
candidate in NGC 55}
\author[A-M. Stobbart et al.]
	{A-M.\ Stobbart, T.P.\ Roberts, R.S.\ Warwick\\  
X-ray \& Observational Astronomy Group, Dept. of Physics \& Astronomy, 
University of Leicester, Leicester LE1 7RH, U.K.\\}

\date{Submitted}

\pagerange{\pageref{firstpage}--\pageref{lastpage}}
\pubyear{2003}

\maketitle

\label{firstpage}

\begin{abstract}

\xmmn EPIC observations have revealed a bright point-like X-ray
source in the nearby Magellanic-type galaxy NGC 55.  At the distance
of NGC 55, the maximum observed X-ray luminosity of the source, designated as
XMMU J001528.9-391319, is $L_x \sim 1.6 \times 10^{39}$ erg s$^{-1}$, 
placing the
object in the ultraluminous X-ray source (ULX) regime.  The X-ray
lightcurve exhibits a variety of features including a significant
upward drift over the 60 ks observation.  Most notably a series of
X-ray dips are apparent with individual dips lasting for typically
100--300 \,s. Some of these dips reach almost 100 percent diminution
of the source flux in the 2.0--4.5 \,keV band. The EPIC CCD spectra
can be modelled with two spectral components, a very soft powerlaw
continuum ($\Gamma \approx 4$) dominant below 2 keV, plus a
multi-colour disc (MCD) component with an inner-disc temperature $kT
\approx 0.8$ \,keV.  The observed upperward drift in the X-ray flux
can be attributed to an increase in the level of the MCD component,
whilst the normalisation of the powerlaw continuum remains
unchanged. The dipping episodes correspond to a loss of signal from
both spectral components, although the blocking factor is at least a
factor two higher for the MCD component. XMMU J001528.9-391319 can be
considered as a candidate black-hole binary (BHB) system. A plausible
explanation of the observed temporal and spectral behaviour is that we
view the accretion disc close to edge-on and that, during dips,
orbiting clumps of obscuring material enter our line of sight and
cause significant blocking or scattering of the hard thermal X-rays
emitted from the inner disc.  In contrast, the more extended source of
the soft powerlaw flux is only partially covered by the obscuring
matter during the dips.

\end{abstract}


\section{Introduction}

\chandra and \xmmn provide powerful facilities for studying the X-ray
properties of nearby galaxies.  A focus of recent work in this area
has been the ultraluminous X-ray source (ULX) phenomenon, namely
point-like X-ray sources located outside the nucleus of the galaxy
with X-ray luminosities apparently in excess of $L_x > 10^{39}$ erg
s$^{-1}$ (\citealt{roberts00}; \citealt{colbert02};
\citealt{miller03}). It is entirely plausible that sources with X-ray
luminosities at, or just above, this threshold are mass-transfer
binaries containing a stellar mass black-hole (3--20 M$_{\odot}$)
radiating at close to the Eddington limit. Supporting evidence is
provided by the fact that many ULXs display similar characteristics to
those of established black-hole binaries (BHBs) (\eg
\citealt{makishima00}).  However, the nature of the subset of ULXs
with X-ray luminosities in excess of a few $\times 10^{39}$ erg
s$^{-1}$ is less certain, since these could be systems harbouring
intermediate mass black-holes \citep{colbert99}, radiating
anisotropically \citep{king01} or possessing truly super-Eddington
discs \citep{begelman02}.

It has been estimated that an accreting black-hole, as opposed to a
neutron star, is present in at least 10\% of all bright X-ray binaries
(XRBs)\citep{mcclintock03}.  At the present time there are only 18
dynamically-confirmed stellar-mass BHBs, most of which were discovered
as X-ray novae in our own galaxy \citep{mcclintock03}.  To this sample
we can add a further 20 or more candidate objects which exhibit all
the characteristics of black-hole systems \citep{mcclintock03}.  Using
\chandra and {\it XMM-Newton}, an individual bright binary X-ray
source can be studied out to a distance of about 10 Mpc, hence studies
of nearby galaxies have the potential for greatly extending our
knowledge of luminous XRBs of all types, including black-hole
systems. For example, M31 has been a prime target for recent
observations (\eg \citealt{kong02}; \citealt{shirey01};
\citealt{osborne01}), and at least one good BHB candidate has 
been identified on the basis of its X-ray properties (RX J0042.3+4115;
\citealt{barnard03}).  The other major Local Group galaxy, M33, also
hosts many discrete X-ray sources (\eg \citealt{haberl01}), including
the most luminous persistent X-ray source in the Local Group (M33 X-8;
\citealt*{trinchieri88}). This source is another good black-hole
candidate, with recent \chandra observations revealing characteristics
consistent with accretion onto a $> 5 M_{\odot}$ object
\citep{laparola03}.  Yet a further example is the discovery of an
eclipsing XRB in NGC 253 \citep{pietsch03}.  In the present paper we
discuss the properties of the brightest X-ray source detected in NGC
55. This source sits right on the boundary of the ``normal''
binary/ULX categorisation and, on the basis of its X-ray luminosity
and spectral properties, is most probably a black-hole system.

NGC 55 is a member of the nearby Sculptor group of galaxies, located
in the region of the South Galactic Pole at a distance of 1.78 Mpc
\citep{kara03}. It is morphologically similar to the Large Magellanic
Cloud but viewed edge-on with its bar pointing almost along the line
of sight ($i=90 \dg$, \citealt{tully88}).  NGC 55 has previously been
studied in the X-ray band through \rosat PSPC (\citealt*{read97};
\citealt*{schlegel97}) and HRI \citep{roberts97} observations.  The
PSPC data revealed seven bright point-like X-ray sources coincident
with the galaxy (\citealt{schlegel97}).  A subsequent re-analysis of
the PSPC data, in conjunction with the HRI data, revealed 25 X-ray
point sources coincident with, or in close proximity to, the disc of
the galaxy \citep{roberts97}.  The \rosat observations showed one
particular object, located $\sim7'$ to the east of the main bar
complex, to be several times brighter than any other X-ray source in
the galaxy (Source 7 of \citealt{schlegel97}; Source 6 of
\citealt{read97}; and Source N55-14 of \citealt{roberts97}).  The
\rosat PSPC data further revealed this source to be spectrally soft
(bremsstrahlung temperature $kT \sim 0.8 - 1.0$ \,keV or powerlaw
photon index $\Gamma \sim 3 - 4$) and mildly absorbed (\nh $\sim 2 - 4
\times 10^{21} \atpcm$) with a derived X-ray luminosity of $\sim 7
\times 10^{38} \ergsec$ in the 0.1 -- 2.4 \,keV \rosat band, adopting
a distance of 1.78 Mpc.  Crucially, both long- and short-term
variability were seen, suggesting this is a luminous accretion-powered
XRB.  Here we revisit this source using new, high quality \xmmn
observations to investigate its spectral and temporal behaviour.

\begin{figure*}
\begin{center}
\scalebox{3}{{\includegraphics[width=50mm,angle=0]{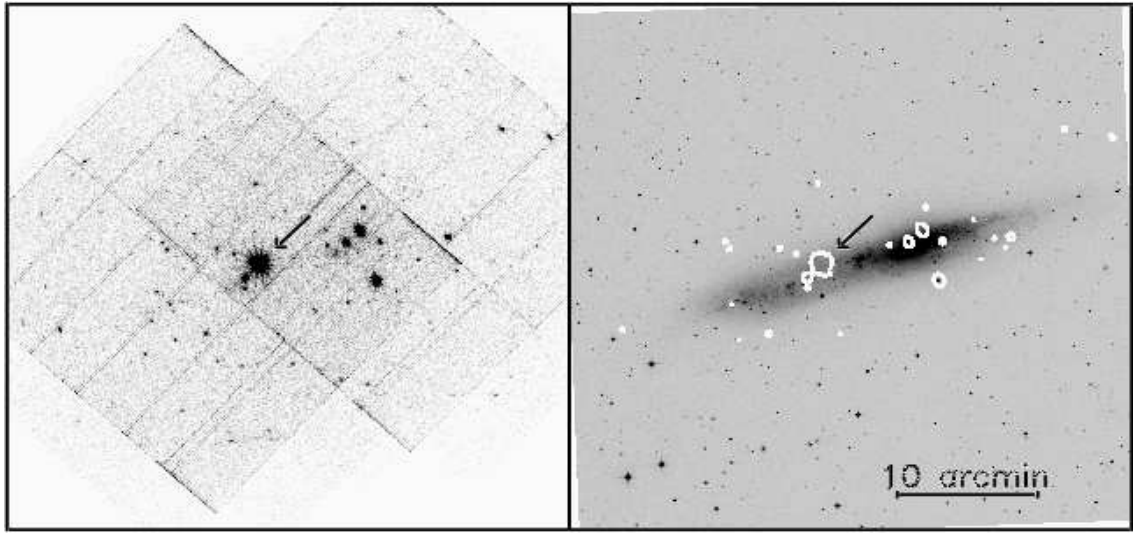}}}
\end{center}
\caption{{\it Left panel:} The \xmmn image of the NGC 55 field in a
broad (0.3--10.0 \,keV) bandpass. The field centre is at RA
$00^{h} 15^{m} 18.0^{s}$, Dec $-39^{\circ} 13' 33''$ (J2000) and 
the image size is $40 \times 40$ arcminute$^2$. The position of
XMMU J001528.9-391319 is highlighted by the arrow.
{\it Right panel:} The equivalent optical DSS-2
(red) image  with the rotation angle aligned precisely to a north-south
projection.  The white contour, which is derived from a lightly 
smoothed version of the X-ray image (using a circular Gaussian 
mask with $\sigma = 1$ pixel $ = 4''$), corresponds to a surface 
brightness of 16 count pixel$^{-1}$. North is up and East is to the left in each case.}
\label{images}
\end{figure*}

\section{The XMM-Newton observations and preliminary data analysis}

NGC 55 was observed during revolution 354 of \xmmn on 2001 November 14
and 15.  This paper focuses on data from the European Photon Imaging
Camera (EPIC, \citealt{turner01}; \citealt{struder01}) taken in the
full window mode and using the thin filter.  Two separate observations
were made, offset from the centre of the galaxy by $\sim$7\arcm in
opposite directions, so as to image the full extent of the galaxy in
the EPIC CCD cameras (see Table 1). The second observation commenced
2.2 ks after the end of the first.  The data were pipeline processed
and reduced using standard tools of {\sc xmm-sas} software v.  5.4.1.
Whilst the instrument background was at a constant low level during
the first observation, the second was affected by soft proton flaring
towards the end of the exposure. A time filter was used to reject the
flaring episode and also to select only those data recorded when all
three cameras (MOS-1, MOS-2 and pn) were in operation. This resulted
in a net exposure time of 30.4 ks and 21.5 ks of data for the first
and second observation respectively with a gap of 6.2 ks between the
two data sets. In the analysis described below we utilise valid pn
events with pattern 0-4 (which have the best energy calibration) but
use pattern 0-12 for the MOS-1 and MOS-2 cameras.

Images were produced in the 0.3--10.0 \,keV energy band for each
observation and for each camera and were then combined using the {\it
emosaic} SAS task. The resulting mosaiced X-ray image is shown in
Fig. \ref{images}. The figure also shows the corresponding DSS-2
optical image of the galaxy with X-ray contours overlaid.  The
brightest source in the earlier \rosat observations retains this
distinction in the \xmmn observations. We determine its position as RA
$00^h15^m28.95^s$, Dec $-39^{\circ}13'19.1''$ (J2000) with an
uncertainty of $\sim 1$\arcs.  Hereafter we employ the source
designation XMMU J001528.9-391319.

\begin{table}
\begin{center}
\caption{The \xmmn observations of NGC 55}
\begin{tabular}{ccccc} 
\hline
Obs ID	&RA$^a$		&Dec$^a$		&Date & UT$_{start}$\\	
\hline
0028740201	&00~15~46.0	&-39~15~28	&2001-11-14&14:20:08\\
0028740101	&00~14~32.9	&-39~10~46	&2001-11-15&00:24:43\\
\hline
\end{tabular}
\end{center}
\begin{tabular}{c}
$^a$ Epoch J2000 co-ordinates.
\end{tabular}
\label{table1}
\end{table}

\section{The X-ray properties of XMMU J001528.9-391319}

\subsection{X-ray lightcurve}

\begin{figure*}
\begin{center}
\scalebox{1.3}{{\includegraphics[width=50mm,angle=270]{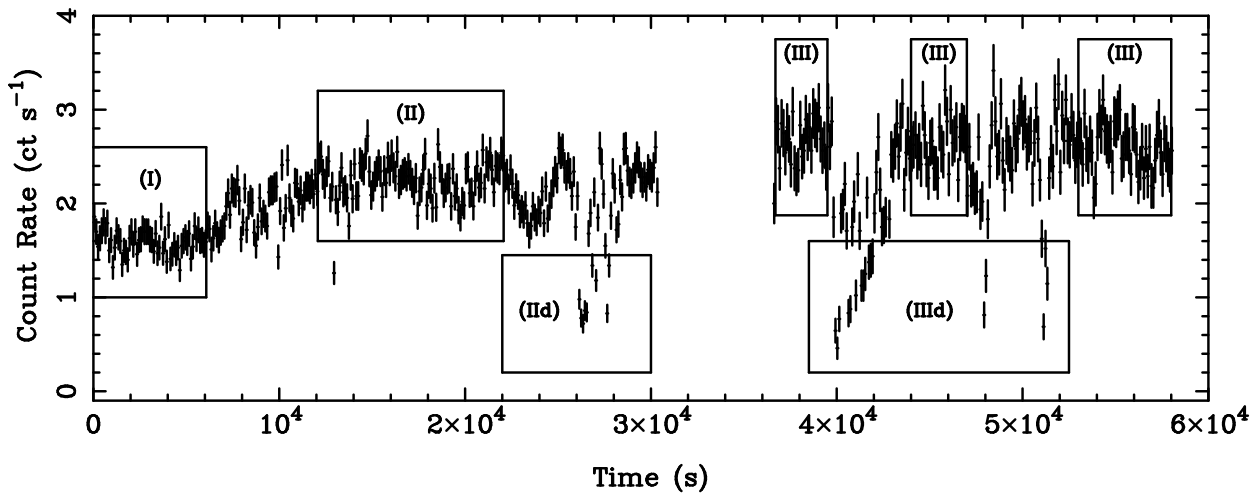}}}
\end{center}
\caption{The background-subtracted lightcurve of XMMU J001528.9-391319 in the 0.3--10.0 \,keV
band in 100 \,s time bins. This lightcurve is based on the combined data
from the MOS-1, MOS-2 and pn cameras.  The different segments
considered in the spectral analysis are identified by the labelled
boxes (see the text for details).}
\label{full_lc}
\end{figure*}

\begin{figure*}
\begin{center}
\scalebox{1.2}{{\includegraphics[width=50mm,angle=270]{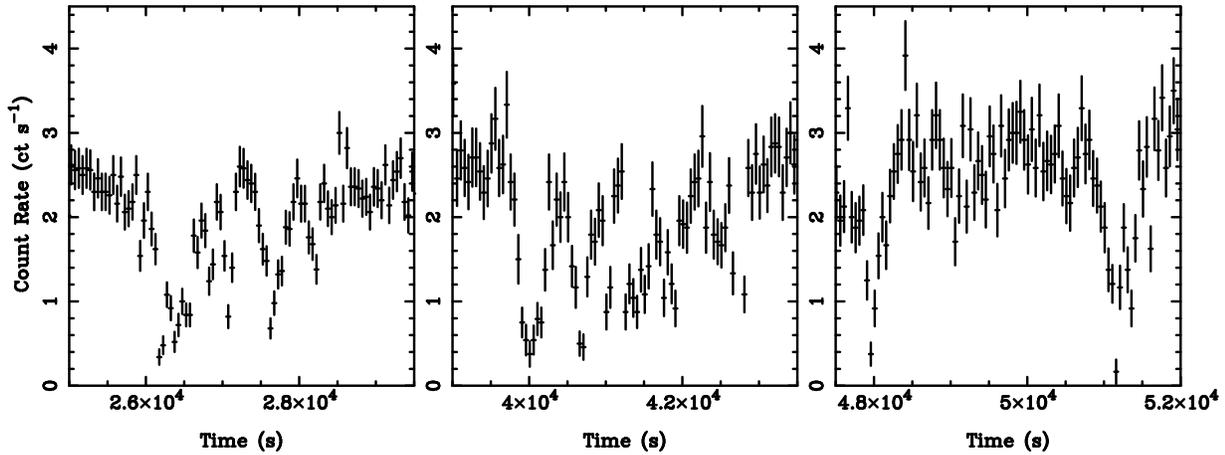}}}
\end{center}
\caption{Zoomed in regions of the 0.3--10.0 \,keV lightcurve of Fig. 
\ref{full_lc}, with 50s time binning. {\it Left panel}: First dipping 
episode. {\it Centre panel}: The first dipping episode of the second 
observation. {\it Right panel}: Remaining dip episodes of the second 
observation.}
\label{dips}
\end{figure*}

A background-subtracted source lightcurve based on the combined data
from the three EPIC cameras was extracted from each observation in the
0.3--10.0 \,keV band. For this purpose we used a source cell of radius
52\arcsec~together with a nearby background region of the same
dimension.  In combining the two datasets we have applied a scaling
factor of 2.1 to the count rates measured in the second exposure, so
as to account for the extra vignetting arising from the large off-axis
angle of the source in this observation\footnote{This correction was
determined via the Ancillary Response Files used in the spectral
analysis.}. The resulting lightcurve is shown in Fig. \ref{full_lc}.
The corrected count rate of the source (summed over the three EPIC
cameras) varies from 0.5--3.5 count s$^{-1}$ with considerable
temporal structure including a marked upperward drift, a significant
level of underlying chaotic activity and pronounced dips.
Fig. \ref{dips} illustrates the structure of the dips in more detail
and shows that a typical dip lasts between 100--300 \,s and that some
of the dips correspond to a 80--90 \% reduction in the source flux in
the 0.3--10.0 \,keV band.

\begin{figure*}
\begin{center}
\scalebox{1.2}{{\includegraphics[width=100mm,angle=270]{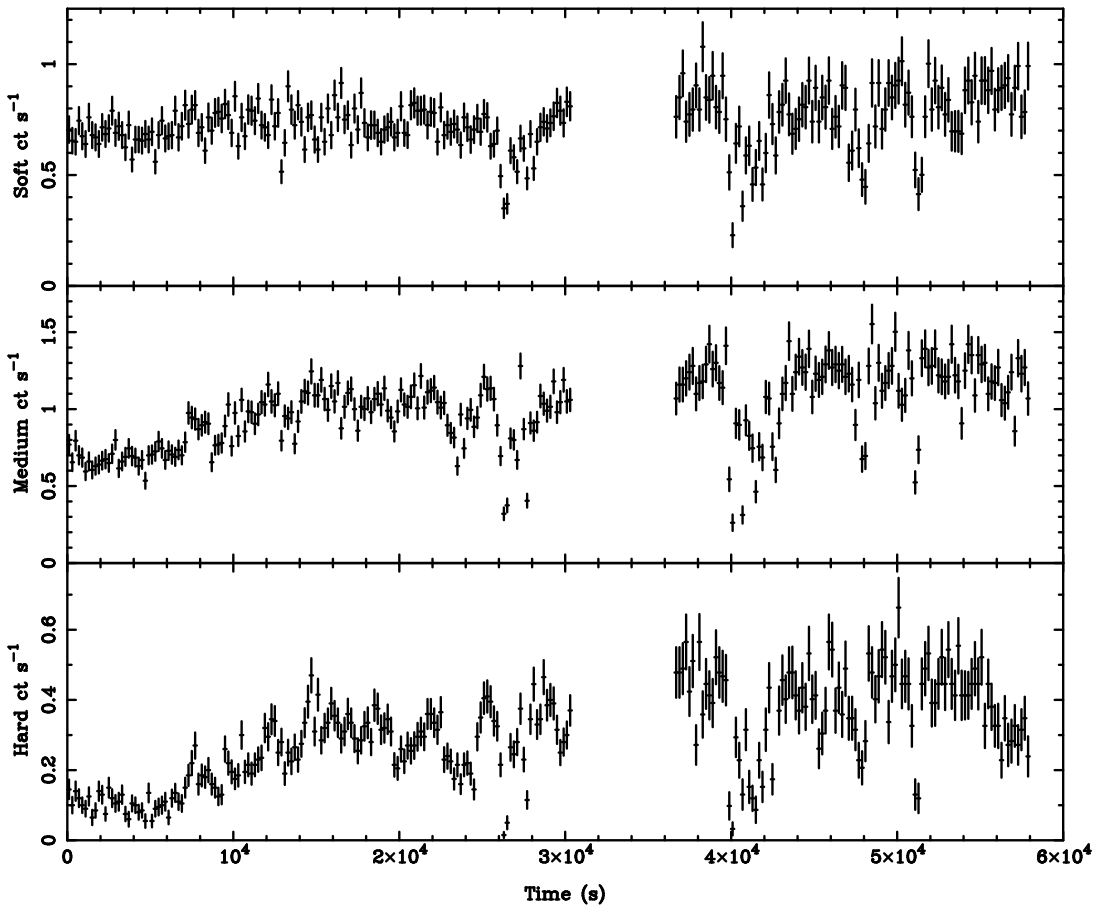}}}
\end{center}
\caption{The lightcurve of XMMU J001528.9-391319 in three energy bands.
From top to bottom: 0.5--1.0 \,keV (soft), 1.0--2.0 \,keV (medium) and 
2.0--4.5\,keV (hard). The time binning is 200 \,s.}
\label{3band_lc}
\end{figure*}

The possible spectral variability of the source was next investigated
by extracting lightcurves, using the method described above, in three
energy bands which optimised the signal to noise ratio: 0.5--1.0 \,keV
(soft), 1.0--2.0 \,keV (medium) and 2.0--4.5 \,keV (hard).
Fig. \ref{3band_lc} shows that the upward drift apparent in the
broad-band lightcurve throughout the observation is due largely to
changes above 1.0 \,keV.  The dips are present in all three bands
although the depth of the dips appears to be somewhat greater in the
harder bands, such that the maximum flux diminution approaches 100\%
in the 2.0--4.5 \,keV band.

\subsection{X-ray spectra}

%201 PN high and low spectra, and model plot 
\begin{figure*}
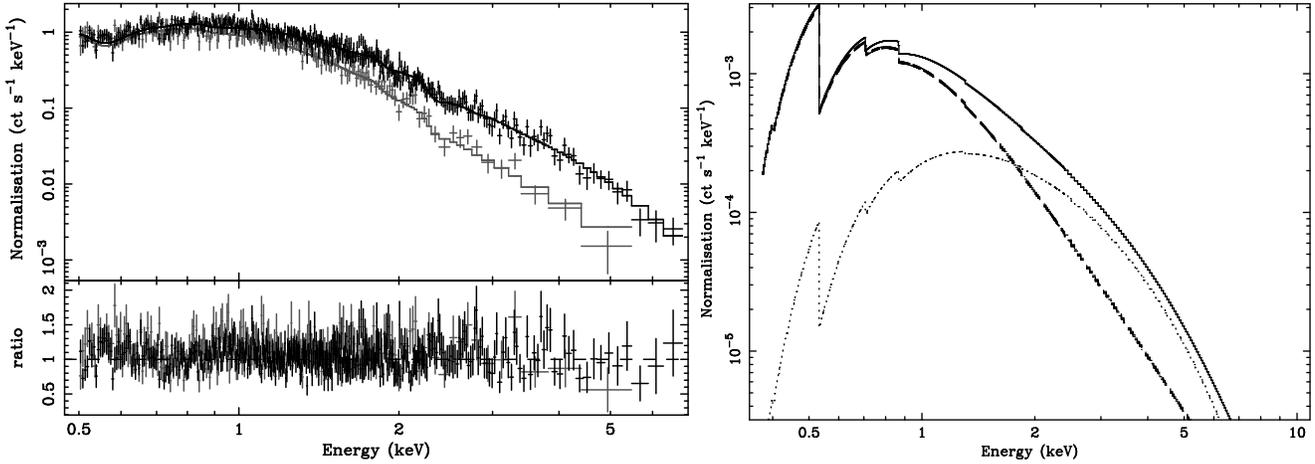

\begin{center}
\includegraphics[width=61mm,angle=270]{figure5a.ps}
%\hspace*{1.0cm}
\includegraphics[width=61mm,angle=270]{figure5b.ps}
\end{center}
\caption{{\it Left panel}: EPIC pn count rate spectra for states I and II
plus the corresponding best-fitting models. {\it Right panel}: The
spectral model used to fit the measured count rate spectra. The model
comprises a steep powerlaw component (dashed line) plus a 
MCD component (dotted line). State I requires only the former component
whereas state II is best-fitted by the combination of the two (solid line).}
\label{201}
\end{figure*}

\begin{figure}
\includegraphics[width=80mm,angle=0]{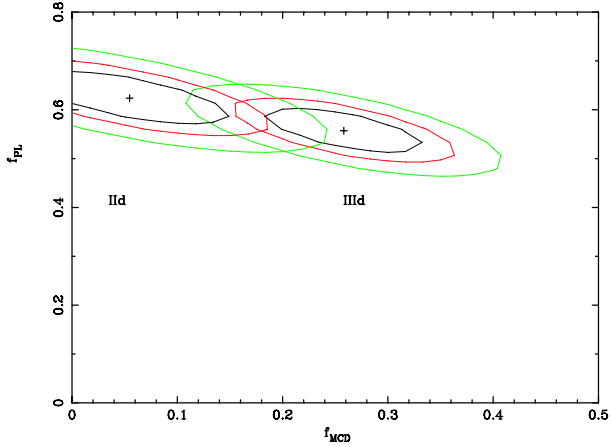}
\caption{Confidence contours (68\%, 90\% and 98\% confidence) for the relative
normalisations of the powerlaw continuum and MCD component during
the two dip states.}
\label{contour}
\end{figure}

The different data segments\footnote{The data selection was carried
out using Good Time Interval (GTI) files based on either time or
count-rate criteria.}  used for spectral extraction are illustrated in
Fig. \ref{full_lc}. These represent different flux states of the
source as follows: (I) the initial low, soft spectral state; (II) a
relatively steady part of the lightcurve when the flux was at an
intermediate level; (III) the highest state reached by the source in
the present observations; (IId) and (IIId), the dipping states
observed in the first and second exposures respectively.

X-ray spectra were extracted for the different data segments using
circular source and background regions of 75\arcs and 110\arcs radii
respectively.  The relevant SAS task {\it especget} produces a source
and background spectrum together with the Ancillary Response File
(ARF) and Redistribution Matrix File (RMF) required in the spectral
fitting.

The spectral analysis was performed using the HEAsoft X-ray spectral
fitting package {\sc xspec} (v. 11.0.1).  Spectral channels were grouped 
so as to give a minimum of 20 counts
per bin, thus ensuring \chisq statistics are valid.  The pn, MOS-1 and
MOS-2 spectra for each data segment were fitted simultaneously, but
with free relative normalisations so as to correct for slight
calibration differences between the cameras.  Spectra were initially
extracted in the 0.3--10.0 \,keV band.  However, since very few counts
were recorded above 7.0 \,keV and there were possible calibration
uncertainties below 0.5 \,keV, the spectral fitting was restricted to
the 0.5--7.0 \,keV range.

As a preliminary step, the spectra recorded while the source was in
its initial low soft state (state I) were fitted with a simple
absorbed powerlaw model.  This resulted in a good match to the data
($\chi^2 = 341$ for 346 degrees of freedom) with parameter values \nh
$\approx 4 \times 10^{21}$ cm$^{-2}$ and $\Gamma \approx 4$. The
former is considerably higher than the line-of-sight column density
through our Galaxy (\nh$ = 1.55 \times 10^{20}$ cm$^{-2}$;
\citealt{stark92}) and may represent the foreground column density
within NGC 55 or material intrinsic to the source itself.  We also tried
a variety of other single-component continuum models but none of these
improved upon the powerlaw fit; for example a single-temperature blackbody 
component yielded $\chi^2 = 459$, whereas a 
MCD spectrum gave $\chi^2 = 400$ (for 346 d-o-f in both cases). 
We also tried the combination of a soft powerlaw
plus a MCD component, but we found that the latter component 
contributed only $\sim$10\% of the total X-ray flux and the
resulting improvement in $\chi^2$ was not significant ($\chi^2 = 338$ 
for 344 d-o-f).

%Table2 fit parameters
\begin{table*}
\begin{center}
\caption{Spectral fitting results for XMMU J001528.9-391319}
\begin{tabular}{ccccccccc} 
\hline
State	&\nh$^{a}$  		&A$_{PL}$ $^{b}$	&$\Gamma$		&A$_{MCD}$ $^{c}$  	&$kT_{in}$$^{d}$  	&\chisq /dof 	&$f_X$$^{e}$	&$L_X$$^{f}$ \\
\hline
Best-fitting model\\
I	&4.2$\pm{0.2}$		&3.0$\pm{0.2}$ 		&4.2$\pm{0.1}$ 		&0.00$+0.10$		&0.86$\pm{0.02}$	&1615/1436	&2.2		&0.85\\
II	&    ''			&  ''  			& ''          		&0.24$\pm{0.03}$  	&  ''			& ''		&3.9		&1.46\\
III	&    ''			&  ''  			& ''          		&0.29$\pm{0.04}$  	&  ''   		& ''		&4.2		&1.59\\
\hline
Standard model\\
I   &1.6$^{+0.2}_{-0.1}$	&0.05$^{+0.06}_{-0.03}$	&2.0$\pm{0.1}$		&7.21$^{+1.51}_{-1.02}$	& 0.41$^{+0.01}_{-0.02}$&2173/1436	&2.2   		&0.81\\
II  &    ''			&0.56$^{+0.10}_{-0.05}$	& ''          		& ''  			&   ''			& ''		&4.2  		&1.57\\
III &    ''			&0.67$^{+0.11}_{-0.05}$	& ''          		& ''  			&   ''   		& ''		&4.6		&1.75\\

\hline
&	&        &            &           &           &    &   & \\
\multicolumn{9}{l}{$^{a}$  Column density ($10^{21}$ cm$^{-2}$).}\\
\multicolumn{9}{l}{$^{b}$  Powerlaw normalisation ($10^{-3}$ phot~cm$^{-2}$~s$^{-1}$).}\\
\multicolumn{9}{l}{$^{c}$  MCD normalisation (${(R_{in}/km)/(D/10kpc)}^2 \cos i$, where $R_{in}$ is the inner disc radius,D the distance to the source and i}\\
\multicolumn{9}{l}{\hspace{0.17cm}	   the inclination angle of the disc).}\\
\multicolumn{9}{l}{$^{d}$  MCD inner-disc temperature (keV).}\\
\multicolumn{9}{l}{$^{e}$  Observed 0.5--10.0 \,keV X-ray flux ($10^{-12}$ erg cm$^{-2}$ s$^{-1}$).}\\
\multicolumn{9}{l}{$^{f}$  Derived 0.5--10.0 \,keV X-ray luminosity ($10^{39}$ erg s$^{-1}$).}\\
\end{tabular}
\end{center}
\end{table*}

Given the evidence from the lightcurves of the presence of a relatively 
steady soft component together with a more variable hard component, we next
attempted to fit the intermediate and high states of the source
(states II and III respectively) with a two component model comprising
the soft powerlaw plus an additional hard component.  In the event, 
trials showed that a MCD model provided the best match to the spectral
shape of the hard emission. The full analysis was conducted by fitting 
the spectra for states I, II and III {\it simultaneously} with the 
normalisation of the powerlaw continuum, the slope of the powerlaw 
continuum, the inner disc temperature $T_{in}$ of the MCD component 
and the column density (applied to both spectral components) tied across 
the three states. The normalisation of the MCD component for each state 
provided the remaining free parameters of the model.  The results of 
fitting this two-component model are summarised in Table 2, where the errors 
are quoted at the 90\% confidence level for one interesting
parameter. This prescription resulted in a reasonable fit to the three
spectral datasets (\chisq of 1615 for 1436 d-o-f).  Fig. \ref{201}
illustrates the changing form of the spectrum between states I and II,
due to the increase in the MCD component.

The combination of a MCD component with a powerlaw is the spectral model 
generally employed for BHB systems (\citealt{mcclintock03}). However,
in the ``standard'' BHB spectral model the powerlaw invariably
represents the hard tail of the emission rather the softest emission
(as above). In essence our best fit model has reversed the role
of the two components compared to the standard picture. 
As a check of our procedure, we have attempted to fit the spectra for the 
three states with the powerlaw component representing
{\it the harder emission} in line with the standard model. Paralleling the 
previous approach, the normalisation of the MCD component, the inner disc 
temperature $T_{in}$, the slope of the powerlaw continuum and the column 
density (applied to both spectral components) were tied across the 
three states with the normalisation of the powerlaw component for each state 
providing the remaining free parameters of the model. The results, which are
summarised in Table 2 under the heading ``standard model'', confirm
our earlier conclusion, namely that the standard description
is not the preferred option in this case.  In our best fitting
model the MCD component provides a good match to the curvature of the 
hard spectrum whereas the soft emission appears to have a simple powerlaw 
form (after allowing for modest excess absorption).

The total source luminosity (excluding the dips) increases by a factor 
$\sim 2$ over the course of the 60 ks interval of the observation, from
$8.5 \times 10^{38}$ erg s$^{-1}$ in the early stages to a maximum of
$1.6 \times 10^{39}$ erg s$^{-1}$ during the latter part of the
observation (Table 2).  Here the luminosities are quoted for the broad 
0.5--10.0 \,keV band and on the basis of a distance to NGC 55 of 1.78 Mpc.

Finally the spectral properties of the dips were investigated by
fitting the spectra from states IId and IIId.  Since the lightcurve
analysis indicates that the relative depth of the dips increases with
energy, this immediately rules out models in which the dipping is the
result of obscuring material entering the line of sight so as to fully
cover both spectral components simultaneously. This is true for either
photoelectric absorption or pure electron scattering.  We can,
however, model the energy dependence of the dipping simply by assuming
different ``blocking factors'' for the two spectral components.  For
each dip state, the parameter values in the model were fixed at the
best-fitting values obtained for the corresponding non-dip data (see
Table 2). Two additional parameters, namely the fractional
normalisations of the soft powerlaw and MCD components during the
dips, $f_{PL}$ and $f_{MCD}$ respectively, were then used to fit the
dip data. Excellent fits to the observed dip spectra were obtained by
this process. Fig. \ref{contour} shows the variation in $\chi^{2}$ as
a function of the two parameters.  For state IId the average flux from
the powerlaw component recorded during the dip intervals was $\sim
60\%$ of the non-dip value but the signal loss for the MCD component
was a factor $>3$ higher. Similarly for state IIId, the blocking of
the MCD component is roughly twice that of the powerlaw continuum.
Unfortunately the dip spectra are of insufficient quality for the
spectral dependence of the blocking factors to be investigated.

\section{Discussion}

\subsection{XMMU J001528.9-391319 as a BHB candidate}
\begin{figure}
\begin{center}
\scalebox{1.2}{{\includegraphics[width=60mm,angle=270]{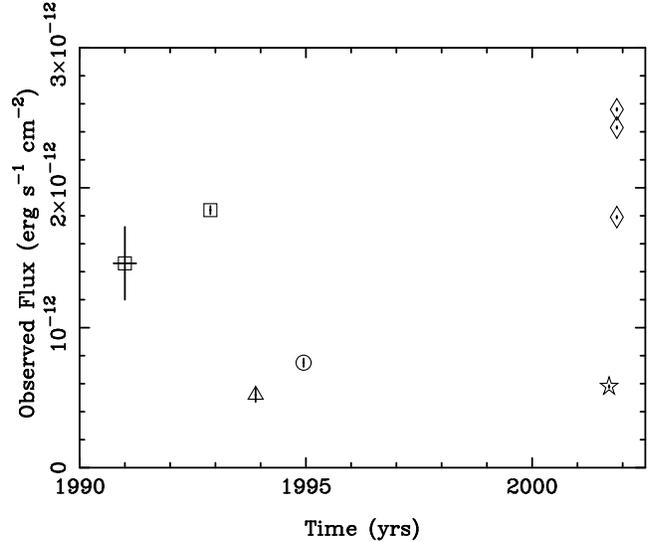}}}
\end{center}
\caption{Long term variation of XMMU J001528.9-391319. Flux measurements 
shown are from \rosat PSPC (squares), \asca (triangle), \rosat HRI (circle), 
\chandra (star) and our three \xmm states (diamonds), in the 0.5--2.0 \,keV 
band. All data were retrieved from archival sources using the standard 
techniques.}
\label{ltlcurve}
\end{figure}

Measurements from ROSAT (PSPC and HRI), {\it ASCA}, \chandra and \xmmn
spanning $\sim10$ years confirm that XMMU J001528.9-391319 is a
persistent highly luminous source (see Fig. \ref{ltlcurve}). The
maximum X-ray luminosity observed by \xmmn of $1.6 \times 10^{39}$ erg
s$^{-1}$ (0.5--10.0 \,keV) places the source at the lower boundary of
the ULX regime.  The variability evident on both long and short
timescales is consistent with XMMU J001528.9-391319 being an
accretion-powered XRB.

If the source is radiating isotropically at or below the Eddington
luminosity, then the implied mass of the compact object is $M > 11$
$\rm M_{\odot}$. This compares to the 3 - 18 \Msun range inferred for
the primary object in the 18 BHBs in our Galaxy and the LMC for which
dynamical measurements are available \citep{mcclintock03}.  As noted
by \citet{mcclintock03}, the three BHBs classed as persistent sources
(Cyg X-1, LMC X-1 and LMC X-3) have high mass secondaries, but
generally radiate well below their Eddington limit.  The other 15 out
of the 18 confirmed BHBs are X-ray novae, and have low-mass
secondaries. Three of these novae have maximum observed X-ray
luminosities firmly in the ULX regime and were ``super-Eddington'' at
peak flux, but of these three only GRS1915+105 appears similar to XMMU
J001528.9-391319, having also been in outburst for more than ten
years.  However, since the neutron star binary A0535-668 achieved a
super-Eddington luminosity at its peak of $L_x \approx 10^{39}$ erg
s$^{-1}$ \citep{white78} it would be unwise to rule out neutron star
models for XMMU J001528.9-391319 on energetic grounds alone.

In order to address the underlying nature of XMMU J001528.9-391319 we
can ask whether the source shows the spectral characteristics of a
stellar-mass BHB.  It certainly displays no evidence of the cool
($kT_{in} \sim 0.1 - 0.2$ \,keV) MCD component seen in more luminous
ULXs.  In fact, the derived 0.86 \,keV inner-disc temperature is
firmly in the observed regime ($kT_{in} = 0.7 - 2.0$ \,keV) for
Galactic black-hole systems in either the thermal-dominated
(high/soft) state or the (very high) steep powerlaw state described by
\citet{mcclintock03}.  Furthermore, the $\Gamma = 4.2$ steep powerlaw
slope is also consistent with either of these states (for which
typically $\Gamma > 2.4$).  We can extrapolate our best-fitting models
to find the flux balance between the two spectral components in the 2
-- 20 \,keV range and use this to try to distinguish between the two
states as per \citet{mcclintock03}.  At the start of the observation
$100\%$ of the flux arises in the powerlaw continuum (using our adopted model 
of a single powerlaw fit to the low state data, but see earlier), falling to
$>27\%$ by the end of the observation.  This points to the very-high
steep powerlaw state, but the detection of a QPO in the 0.1-30.0 Hz
range is really required in order to completely resolve the
ambiguity. This very-high state has frequently been associated with
the highest accretion rates in Galactic BHBs, consistent with the high
luminosity of XMMU J001528.9-391319.

However, there is a potential problem with this interpretation.  In
XMMU J001528.9-391319 we see the spectral components inverted with
respect to the standard spectral description of Galactic BHBs, {\it i.e.,} 
the soft end of our
spectrum is dominated by the powerlaw continuum, not the MCD
component.  To the best of our knowledge this is the first example
of a BHB candidate in which the powerlaw continuum ``re-emerges'' at low
energies ({i.e.,} below 2 keV) as the MCD continuum turns down.  
Such behaviour has not been seen previously even in sources
with a relatively low absorption column, similar to the $\sim 4 \times
10^{21} \atpcm$ determined for XMMU J001528.9-391319; for example, see
the recent observations of XTE J1550-564 \citep{millerj03} and GX 339-4
\citep{millerj04}.  On the other hand, there are BHBs with higher
absorption columns that possess observed spectra in the
thermal-dominated state that, if extrapolated below 2 keV, would be
dominated by an emergent powerlaw component (e.g. XTE J1550-564 and
GRS 1915-105; \citealt{mcclintock03}). Furthermore some BHBs in the steep
powerlaw state have spectra dominated by their powerlaw component at
all observed energies (e.g. GRO J1655-40).  XMMU J001528.9-391319
appears to be a source in which the powerlaw and MCD components are
finely balanced in terms of which happens to dominate in any particular 
spectral regime.

However, the spectrum of XMMU J001528.9-391319 could
pose a challenge to the current paradigm for the X-ray emission from
BHBs, which has the powerlaw continuum originating in an extended
Compton-upscattering medium (the corona) above the accretion disc.
This is because an extrapolation of the powerlaw continuum to
very soft X-ray energies may be problematic due to a lack of soft seed
photons from the accretion disc at energies below the turnover in  the 
MCD continuum.  To solve this, one might invoke either a new source of 
seed photons or recourse to a different process for the production
of the soft continuum.  For example, synchrotron radiation from
the innermost part of a jet, has been discussed as a possible origin
of the powerlaw component in some BHBs (e.g. \citealt{markoff01}),
although jets may be quenched for BHBs in a high-soft or very-high 
state (e.g., \citealt{fender04}).

A second estimate of the mass of the compact object can be obtained
from the parameters of the MCD model. On the basis of the spectral
analysis we determine the peak bolometric flux of the disc component
to be $3.4 \times 10^{-12} \ergcms$ and using equation (5) of
\citet{makishima00}, we obtain an inner-disc radius of $R_{in} = 113$
(cos $i$)$^{-{{1}\over{2}}}$ km, where $i$ is the inclination of the
accretion disc to the line of sight.  For a Schwarzschild black-hole
this converts to a mass of $12.8$ (cos $i$)$^{-{{1}\over{2}}}$ $\rm
M_{\odot}$ (equation 8, \citealt{makishima00}).  In Table 3 we compare
the masses derived via the MCD model with the known dynamical mass for
five BHBs.  We may interpret this tabulation either in terms of 3 out
of 5 of these systems having very high inclinations ($i > 80\deg$) or
the above method tending, in some cases, to underestimate the true
black-hole mass by a factor of several (which might indicate that some
of these black-holes carry significant angular momentum,
\citealt{makishima00}). Interestingly, in the limit cos $i = 1$, the
MCD method never {\it overestimates} the mass of the black hole.
Since our system is dipping this gives us a crude handle on the
inclination of its accretion disc. If we conservatively assume $i \sim
70\deg$ (see below), this leads to a mass estimate of $ M_{MCD} \ga
20$ \Msun, implying that this system may harbour a black-hole more
massive than in any confirmed BHB.

\begin{table}
\begin{center}
\caption{Mass estimates from the MCD model compared to dynamical
measurements.$^a$}
\begin{tabular}{lcc} \hline
System & $M_{MCD}$ (\Msun) & $M_{dyn} (\Msun)$
\\
\hline
4U 1543-475	& 8.5  (cos $i$)$^{-{{1}\over{2}}}$  & 7.4 -- 11.4 \\
XTE J1550-564	& 3.4   (cos $i$)$^{-{{1}\over{2}}}$ & 8.4 -- 10.8 \\
GRO J1655-40	& 1.7  (cos $i$)$^{-{{1}\over{2}}}$  & 6.0 -- 6.6 \\
GX 339-4	& 2.7  (cos $i$)$^{-{{1}\over{2}}}$  & $\sim 5.8$ \\
GRS 1915+105	& 1.2  (cos $i$)$^{-{{1}\over{2}}}$  & 10 -- 18 \\
\hline
\end{tabular}
\end{center}
\begin{tabular}{c}
$^a$ Based on data presented in McClintock \& Remillard (2003)
\end{tabular}
\label{table3}
\end{table}

\subsection{XMMU J001528.9-391319 as a dipping BHB candidate}
The most interesting feature of XMMU J001528.9-391319 is the dipping
apparent in its lightcurve. Such activity has been reported in $\sim
20$ XRBs including at least two BHBs, \eg GRO J1655-40
(\citealt{oroz97}) and Cyg X-1 (\citealt{kitamoto84}). To the best of
our knowledge, XMMU J001528.9-391319 is the first extragalactic
black-hole candidate seen to exhibit dip behaviour.

A variety of models have been put forward to explain dipping activity
with the details dependent on whether the system in question is a
low-mass XRB (LMXRB) or high-mass XRB (HMXRB).  In LMXRBs, dipping is
thought to be due to the obscuration of the central X-ray source by
absorbing material in the region where the accretion flow from the
companion star impacts on the outer accretion disc of the system
(\citealt{white82}; \citealt{white95}).  This could be via an
interaction of the gas stream from the secondary with the outer edge
of the accretion disc, causing a thickening at the rim (\eg EXO
0748-676; \citealt{parmar86}). Alternatively the obscuration may arise
not from a bulge on the accretion disc itself but from a filamentary
or clumpy absorbing medium extending above or below the impact region
(\eg GRO J1655-40; \citealt{kuulkers98b}).  Often in these cases the
dips are observed at certain orbital phases and most of these sources
are described as periodic dippers (\citealt{white82}). In contrast, in
HMXRBs the absorbing material is generally associated with clouds or
``blobs'' in the stellar wind of the companion star. In these cases
there are often sharp transitions into and out of the dip states,
accompanied by variations in the column density, giving evidence for
nonuniformity (\ie clumpy material) in the stellar wind.  In the case
of Cyg X-1 dips occur preferentially at orbital phase $\phi \sim 0.95$
with a secondary peak at $\phi \sim 0.6$, features which have been
interpreted respectively in terms of the influence of the stellar wind
and of an accretion stream (\citealt{kitamoto84}; \citealt{church00}).
Although the X-ray flux from Cyg X-1 significantly photoionizes the
stellar wind of its supergiant companion star, HDE 226868, the factor
100--1000 overdensity of the clumps which give rise to the observed
dips, explains the near-neutral characteristics of the absorption in
this source (\citealt{kitamoto84};
\citealt{church00}).

Since XRBs which show dips but not eclipses typically have
inclinations in the range $60\deg - 75\deg$ (\citealt{frank87}), a
relatively edge-on configuration seems very likely for XMMU
J001528.9-391319.  The short 100--300 \,s duration of the dips mirrors
behaviour seen in the BHBs GRO J1655-40 (\citealt{kuulkers00}) and Cyg
X-1 (\citealt{kitamoto84}) which are respectively, low-mass and
high-mass binary systems.  This short timescale constrains the
dimensions of the obscuring clumps. For example, if these clumps are
orbiting a 20 \Msun black-hole, in or just above its accretion disc,
with a typical clump encompassing less than 1\% of the disc
circumference, then the size of a typical clump must be $ \la 10^{10}$
cm at a radius of $ \ga 2 \times 10^{11}$ cm.  For unity optical depth
to electron scattering the required density is $ \ga 1.5 \times
10^{14}$ cm$^{-3}$ at which point the ionization parameter $ \xi =
L/nR^{2} \la 200$ and the material may be quite strongly photoionized
(\citealt{hatchett76}; \citealt{kuulkers98a}). The orbital timescale
of such a clump is $\sim 10000$ s. Although several dip episodes are
apparent over a timescale of $\sim 20$ ks, within the bounds of our
limited observations, it is not possible to investigate whether the
dips follow a periodic pattern.

Our spectral analysis suggests that the MCD component is more strongly
obscured than the powerlaw component during the dips, but we have no
constraints on whether the obscuration arises from
absorption/scattering (in effect complete blocking) in dense cold
clumps or is due to electron scattering in a much hotter medium.  A
scenario which matches the limited information is one in which the
harder thermal disc emission from the inner parts of the accretion
disc is almost completely blocked during a dipping episode whereas the
more extended softer emission (possibly from a hot corona above the
disc, though see the discussion above) is only partially obscured by the
same clump.  However, there are other plausible explanations of the
apparent spectral softening during dips. For example photoionization
may have significantly reduced the soft X-ray opacity of the obscuring
medium or alternatively the scattering of X-rays in hot clouds above
the accretion disc might induce an apparent soft excess during dips
\citep{frank87}. Unfortunately the quality of the X-ray spectra
recorded during the dips is too poor for us to be able to distinguish
between these possibilities.

\section{Conclusions}

\xmmn observations have revealed a luminous XRB in the nearby galaxy
NGC 55. On the basis of its X-ray luminosity and X-ray spectral
properties we believe this object is most likely a black-hole system.
The lightcurve reveals very interesting spectral variability
including, most notably, pronounced dips. Future observations may
reveal whether these dips show a pattern of occurrence consistent with
an underlying orbital period as is the case for the well-studied BHBs
GRO J1655-40 and Cyg X-1.  However, detailed investigation of the
spectral variations which accompany the dips represents a challenge
even for the \xmmn instrumentation.
\section*{Acknowledgments}

AMS and TPR gratefully acknowledge funding from PPARC. This work is
based on observations obtained with \xmmn, an ESA Science Mission with
instruments and contributions directly funded by ESA member states and
the USA (NASA).  The other X-ray data used in this work were obtained
from the Leicester Database and Archive Service (LEDAS) at the
Department of Physics and Astronomy, University of Leicester. The
second Digitized Sky Survey was produced by the Space Telescope
Science Institute, under Contract No. NAS 5-26555 with the National
Aeronautics and Space Administration. We are grateful to the anonymous
referee for very useful comments on this paper.

\label{lastpage}

{}

\end{document}